**Article type: Full Paper**

# Manipulation of Orbital Angular Momentum Spectrum Using Shape-Tailored Metasurfaces

*Ling-Jun Yang, Sheng Sun\*, and Wei E. I. Sha\**

L.-J. Yang, Prof. S. Sun
School of Electronic Science and Engineering
University of Electronic Science and Technology of China
Chengdu 611731, China
E-mail: sunsheng@ieee.org

Prof. W. E. I. Sha
College of Information Science & Electronic Engineering
Zhejiang University
Hangzhou 310027, China
E-mail: weisha@zju.edu.cn

Vortex beams carrying orbital angular momentum (OAM) have been widely applied in various electromagnetic, optical, and quantum systems. A tailored OAM spectrum composed of several specific modes as expected holds a promise for expanding the degrees of freedom of the systems. However, such a broadband high-purity tailored spectrum is difficult to be achieved by the present devices, where the broadband amplitude manipulation has not been explored yet. In this work, inspired by the envelope-modulation theory, an elegant and universal way to manipulate the OAM spectrum in wide bandwidth is proposed by using a shape-tailored metasurface. Firstly, the rotating meta-atoms on a triangular lattice are proved to have smaller coupling distortion than that on a square lattice, and this behavior is critical for high-purity vortex spectrum generation by the Pancharatnam-Berry-based metasurfaces. Secondly, a universal modulation relation is established between the spatial arrangement of metasurfaces and the generated vortex beams. Finally, the broadband modulated OAM spectra and the comb-like OAM spectra are theoretically and experimentally demonstrated by the shape-tailored metasurfaces. The proposed amplitude-modulation scheme offers a novel concept and engineering route to manipulate the OAM spectrum in wide bandwidth, which could promote the development of OAM-based applications.

## 1. Introduction

Orbital angular momentum (OAM), as one of the fundamental physical quantities with additional degrees of freedom,[1] has drawn considerable research interests. The OAM-carrying waves, known as vortex beams, can theoretically enhance the channel capacity by utilizing the orthogonality of different OAM modes $l$.[2] Therefore, the high-capacity communication, based on OAM multiplexing, was experimentally verified in radio,[3,4] optical,[5–7] and quantum fields. However, OAM communication still suffers from some limitations including misalignment, mode crosstalk, large receiving aperture, and over-quadratic power decay.[8,9] Thus, high-purity vortex beams carrying the OAM spectra[10–12] are always highly demanded in multi-mode hybrid communication as well as its applications including super-resolution imaging,[13] optical tweezers,[14] and radar pulse.[15,16]

During the past two decades, enormous effective approaches have been reported to generate vortex beams, such as holographic plates,[17–19] spiral phase plates,[20,21] antenna arrays,[22,23] as well as metasurfaces.[24–32] In addition, PB (Pancharatnam-Berry)-based metasurfaces were recently reported as a promising approach to fabricate ultrathin and high-efficiency OAM generators,[33–40] some of which can also generate high-performance and broadband vortex beams.[41–43] Nevertheless, most metasurfaces were designed by simply manipulating the phase profiles or distributions of incident waves to generated OAM beams, leaving the OAM spectra not rich and hard to be manipulated. Moreover, all of the metasurfaces mentioned above were constructed as square or disk-shaped geometries, which is not to be an optimum choice for multi-mode OAM generation. More recently, the high-dimensional OAM spectrum, especially if the OAM spectrum is tailorable, has attracted considerable attention.[44–49] Specifically, both the quadratic phase plates and the pinhole plates were theoretically demonstrated to generate vortex beams with a comb-like OAM spectrum.[46] Nonetheless, an approach that has both a deep physical understanding and a universal engineering technique to manipulate the OAM spectrum in wide bandwidth has not been explored yet.

In this work, we propose a shape-tailored metasurface based on the triangular-lattice arrangement to generate the vortex beam with a high-performance and controlled OAM spectrum. Firstly, we analyze the influence of lattice topology on the PB phase when meta-atoms are rotated. Compared with the square lattice, the triangular lattice presents a smaller coupling distortion and can be used to fabricate a broadband metasurface with higher performance. Inspired by the envelope modulation theory in the communication system, we establish a universal modulation relation between the shape of metasurfaces and the corresponding OAM spectra of generated vortex beams. Then, the shape-tailored metasurface is proposed, which can generate vortex beams with a tailored OAM spectrum (See **Figure 1**). Specifically, both the envelope-modulated OAM beams and the OAM-comb beams are generated by the proposed shape-tailored metasurfaces. Next, the spectra of the generated OAM beams are analyzed by the Fourier decomposition, confirming the functions and flexibilities of those proposed metasurfaces. The near-field experimental results also demonstrate that the shape-tailored metasurface could well manipulate the OAM spectrum, which is significant for the OAM related applications. Notice that this approach, providing a broadband amplitude-modulated scheme, generates the desired OAM spectrum and could be also applied to other amplitude-oriented devices.

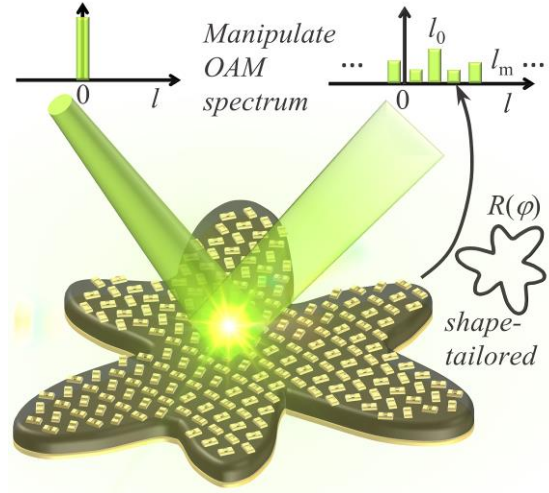

**Figure 1.** A schematic of a shape-tailored metasurface for OAM spectrum modulation.

## 2. Theoretical Design and Discussion
### 2.1. The Distortion in Spin-to-orbital Process

To manipulate the amplitude of the reflection field, it is better to consider all the amplitude-related distortions such as the distortion in the spin-to-orbital process. The spin-to-orbital conversion means the proposed metasurfaces can converge the incident circular polarization plane waves with spin angular momentum into reflected or transmitted vortex waves with orbital angular momentum. The key point is to introduce a spiral abrupt phase by the rotated meta-atoms, which is the so-called geometric or PB phase. The ideal spin-to-orbital process of a reflection-typed meta-atom can be demonstrated as fellow:[50]

$$r_{ll} = 0.5[(r_{xx} - r_{yy}) + j(r_{xy} + r_{yx})]e^{-2j\phi} \quad (1a)$$

$$r_{lr} = 0.5[(r_{xx} + r_{yy}) + j(r_{yx} - r_{xy})] \quad (1b)$$

$$r_{rl} = 0.5[(r_{xx} + r_{yy}) - j(r_{yx} - r_{xy})] \quad (1c)$$

$$r_{rr} = 0.5[(r_{xx} - r_{yy}) - j(r_{xy} + r_{yx})]e^{2j\phi} \quad (1d)$$

where $r_{xx}$, $r_{yy}$, $r_{ll}$, and $r_{rr}$ are the co-polarized reflection coefficients under $x$-, $y$-, left circularly, and right circularly polarized (CP) normal incidences, respectively. $r_{xy}$, $r_{yx}$, $r_{lr}$, and $r_{rl}$ are the cross-polarized reflection coefficients. $\phi$ is the rotation (orientation) angle of the meta-atom. **Figure 2**a shows a traditional square arrangement of the deformed square loop and **Figure 2**b shows the corresponding topology with the rotated meta-atoms. Here, the periodic boundary conditions or the lattice keep unchanged when meta-atoms are rotated. Notice that the rotation of meta-atoms will unavoidably change the coupling relation between adjacent meta-atoms, due to the geometric characteristics of the square lattice. As a result, Equation (1) can work well only if the coupling is small enough to be ignored or at a specific rotation angle of meta-atoms maintaining the rotational symmetry of the lattice. Moreover, it's difficult to eliminate such a coupling especially for the meta-atoms with a broadband operation.[41,42] Alternatively, the arrangement on a triangular lattice has already been applied in phased arrays,[51] Luneburg lens,[52] and high-impedance electromagnetic surfaces[53] to improve their performance. However, very little attention has been paid to the PB-based metasurfaces, where the rotated meta-atoms benefit more from the topology of the triangular lattice. Figure 2d presents the triangular lattice arrangement of meta-atoms. Considering the rotated meta-atoms on different lattice arrangement, a distorted spin-to-orbital process shows as follows:

$$r_{ll}' = r_{ll} + \Delta r_{ll} \tag{2a}$$

$$r_{lr}' = r_{lr} + \Delta r_{lr} \tag{2b}$$

$$r_{rl}' = r_{rl} + \Delta r_{rl} \tag{2c}$$

$$r_{rr}' = r_{rr} + \Delta r_{rr} \tag{2d}$$

where a distorted term $\Delta r$ is added into Equation (1), which is triggered by the change of coupling between meta-atoms when they are rotated. Own to the rotational symmetry of the square and triangular lattice, the distorted term can be expanded as a periodic function.

$$\Delta r = \sum_{n=1}^{\infty}(a_n \cos(nk\phi) + b_n \sin(nk\phi)) \tag{3}$$

where the value of $k$ is taken as 4 and 6, respectively, for the square and triangular lattices. $a_n$ and $b_n$ are complex constant related to coupling. The coupling relation for different lattice can be demonstrated in their equivalent model in Figure 2c,f. Obviously, the triangular lattice arrangement can benefit from the more coupling capacitors (six rather than four) determined by the topology. It means that the triangle arrangement can potentially achieve a lower coupling between each adjacent meta-atom ($C_i(\phi)$) to meet the broadband result. It will also result in a smaller $a_n(b_n)$ or $\Delta r$ than square arrangement case when meta-atoms are rotated. Therefore, the triangular arrangement is a promising candidate to generate wideband high-performance vortex beams.

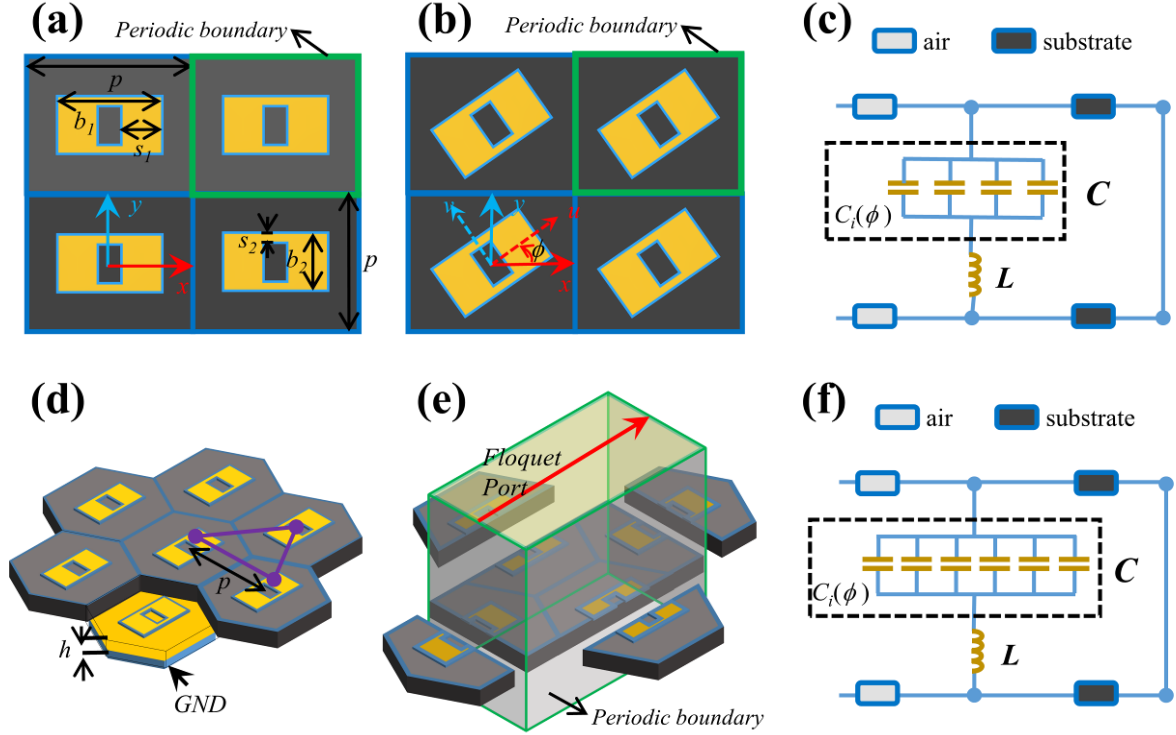

**Figure 2.** The topology of the square arrangement metasurface with a) non-rotated meta-atoms and b) rotated meta-atoms. c) The equivalent transmission-line model of the square arrangement meta-atoms. d) Topology, e) the simulated setup, and f) the equivalent transmission-line model of the triangular arrangement meta-atoms.

## 2.2. The Simulation Results of Arrangement on Triangular Lattice (the hexagonal meta-atoms) and Square Lattice (the square meta-atoms)

Both square-lattice and triangular-lattice arrangements are set to the period between adjacent meta-atoms as $p = 10$ *mm* and designed with a similar bandwidth from 9 GHz to 19 GHz according to the proposed method.[41] In this paper, the design parameters of the square meta-atoms (on square lattice) are $b_1 = 8.3$ *mm*, $s_1 = 3.5$ *mm*, $b_2 = 3$ *mm*, $s_2 = 0.3$ *mm*, $h = 3$ *mm* and $\varepsilon_r = 2.65$; The design parameters of the hexagonal meta-atoms (on triangular lattice) are $b_1 = 7.8$ *mm*, $s_1 = 2.6$ *mm*, $b_2 = 2.6$ *mm*, $s_2 = 0.3$ *mm*, $h = 3$ *mm* and $\varepsilon_r = 2.65$, where some parameters are adjusted to meet the bandwidth due to the difference of coupling structure.

Figure 2e depicts the simulation setup for the hexagonal meta-atoms.[51] By using the commercial software HFSS (High Frequency Structural Simulator), the CP reflection coefficient could be obtained from the linearly polarized field with Floquet port.[36] The results of the square meta-atoms in **Figure 3**a shows that the CP reflection coefficient suffers more from its distortion with respect to the rotation angle of meta-atoms, and this kind of distortion also appears in the previous square broadband meta-atoms.[41,42] However, the hexagonal meta-atoms can significantly reduce the distortion of the reflection coefficient as shown in Figure 3b, therefore a higher expected co-polarized reflection coefficient (0.975) can be obtained (rather than 0.96). Figure 3c shows a detailed variation of reflection coefficient at 11 GHz, where the polarization conversion ratio (PCR) is commonly used to describe the conversion efficiency of the polarization converter and defined as:

$$\text{PCR} = \frac{r_{ll}^2}{(r_{ll}^2 + r_{rl}^2)} \tag{4}$$

The reflection coefficients on different lattice show a four- or six-periodic variation, which is consistent with the distortion Equation (3). The PCR on the triangular lattice is larger than 95.7%, representing a high-efficiency conversion in this hexagonal meta-atom. Moreover, the hexagonal meta-atoms with only 1.3% (rather than the square meta-atoms with up to 5%) fluctuation of PCR is essential for the PB-based metasurface in which the rotation angle of each meta-atom is not the same. Figure 3d,e demonstrate the specific influence on the final performance of the metasurface, where two $l = -2$ metasurfaces have been fabricated by the square and the hexagonal meta-atoms, respectively. Under the LHCP plane wave excitation, the amplitude and phase patterns of far-field $E$-field ($|E|$ and $\angle E$) are depicted in spherical coordinates, where $\varphi$ is the azimuthal angle and $\theta$ is the polar angle. Since the rotation angle $\phi$ of each meta-atom is equal to the azimuthal angle $\varphi$ for the $l = -2$ metasurface, the main lobe of $|E|$ in Figure 3d suffers from a four-periodic variation along the $\varphi$ due to the four-periodic fluctuation of PCR. One can study the specific OAM spectra by implementing Fourier spectrum analyses at their main lobes.[41] The generated OAM purity of $l = -2$ mode by the square atoms is 5% (10%) less than that by the hexagonal meta-atoms at 11 GHz (8.5 GHz). As the low fluctuation of PCR on the triangular lattice, the purity of $l = -2$ mode in OAM spectrum is higher than 96% and crosstalk modes can be ignored in Figure 3e. These simulated results verify that the arrangement on triangular lattice achieves a better performance for broadband PB-based metasurfaces. It is important for other PB-based high-performance applications, and small fluctuation of PCR is also critical for the design of amplitude-related metasurface.

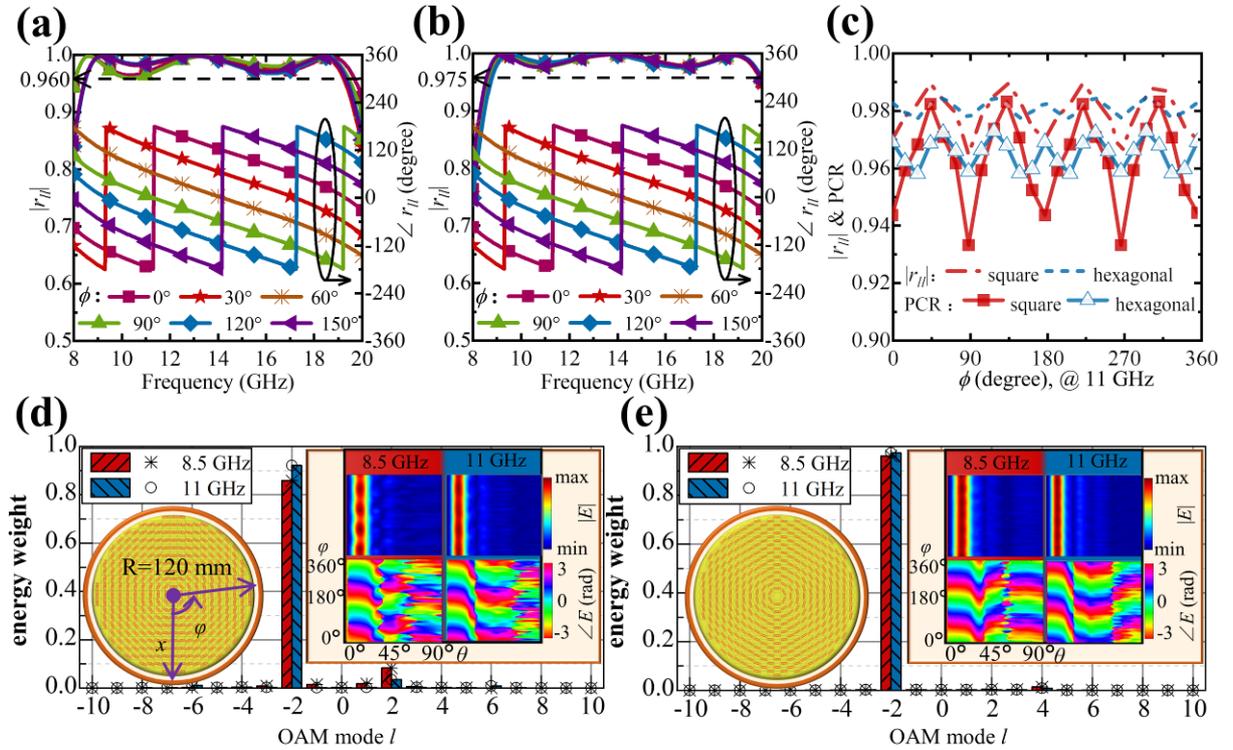

**Figure 3.** The CP reflection spectra for a) the square and b) the hexagonal meta-atoms with different rotation angles. c) The distortion results of the square and the hexagonal meta-atoms at 11 GHz. Far-field co-polarized numerical results under spherical coordinates and corresponding spectral analyses of the $l = -2$ vortex beams generated by d) the square meta-atoms, e) the hexagonal meta-atoms.

## 2.3. The Crosstalk in Specific Shape of Metasurface

In literature, most metasurfaces were arranged in a square shape to fit the square-arrangement cells. However, such a shape led to the nonuniformity of the electric field amplitude, which eventually generated crosstalk modes and affected the purity of the generated vortex wave. **Figure 4** shows the vortex beams results generated by three different shapes of metasurfaces. They are the square-shape, the hexagonal-shape, and the circular-shape metasurfaces with triangular-arrangement meta-atoms. The operating frequency of the excited left-handed circularly polarized (LHCP) plane wave is set to be 13 GHz to avoid the fluctuation of PCR, where the PCR of the meta-atoms at different rotation angles is almost constant. Figure 4a,b show the far-field observation results of the $l = -3$ vortex beam generated by the square-shape and the hexagonal-shape metasurfaces, respectively. The OAM spectral analyses are taken at $\theta = 10°$, which is corresponding to their main lobe in this case. Here, a noticeable periodic fluctuation can be observed in the main lobes of electric field amplitudes, which originates from the modulation effect of the shapes of the metasurfaces. Although three periodic helical phase fronts could be found in the phase distribution, the fluctuation amplitude will result in unwanted co-polarized crosstalk of OAM modes $l = -3 \pm m \times n$ ($n \in Z$) in OAM spectra (where $m$ is 4 or 6 corresponding to the shape of the square or the hexagonal metasurface). It eventually influences the mode purity of the generated vortex wave. As a comparison, an unmodulated uniform amplitude pattern can be produced by a circular metasurface, as shown in Figure 4c. Consequently, a high-purity vortex wave is generated with an energy percentage of $l = -3$ more than 90%, which is 18% higher than that achieved by traditional square metasurface. Hence, the circular-shape metasurface is a better choice for high-purity single-mode vortex beam generation.

Actually, those shape-dependent crosstalks in modulated OAM modes have been partially noticed in the previous works,[54,55] and the corresponding relations between the crosstalk and the rotational symmetry of the meta-atoms have been understood as a general angular momentum conservation law.[56–58] However, an effective method to control and redistribute energy of these modes has not been well studied.

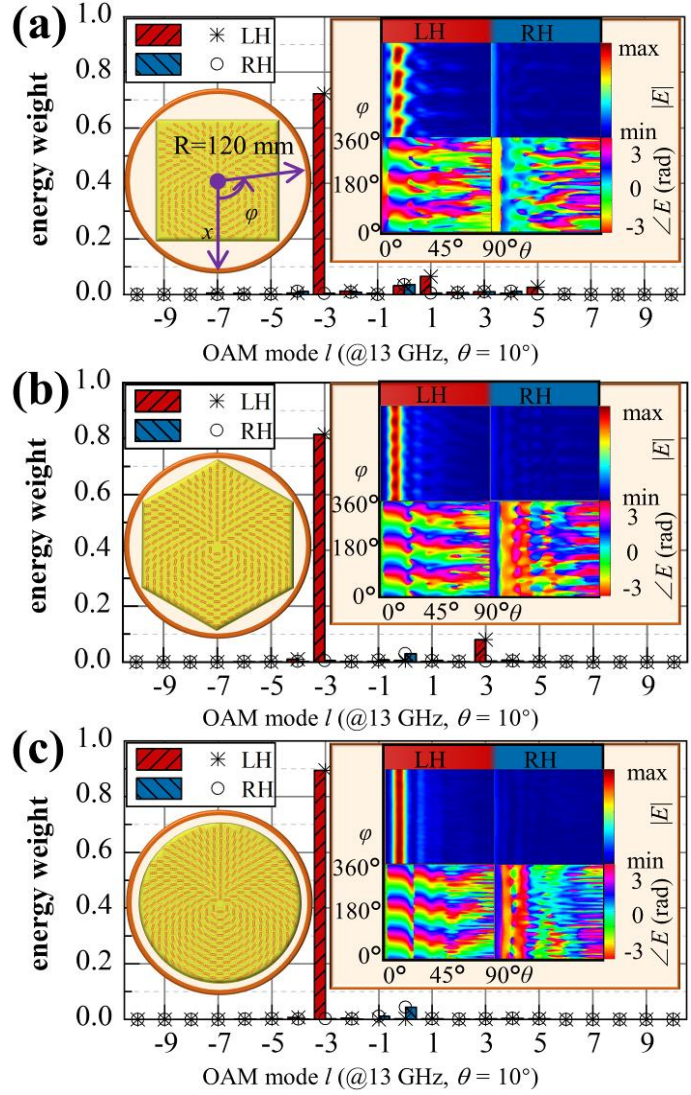

**Figure 4.** Far-field observation and corresponding spectral analyses of vortex beams with the OAM mode $l = -3$ generated by a) the square-shape, b) the hexagonal-shape, and c) the circular-shape metasurfaces under the excitation of the LHCP-plane wave at 13 GHz.

### 2.4. Generation of modulated OAM Spectra

It is well known that amplitude modulation[59] is a widely-used modulation technique used in the communication system. Inspired by the same spirit, we could write down an envelope-modulated OAM vortex beam as follows:

$$\psi(\varphi) = (1 + \sum_i a_i \sin(l_i \varphi)) A_{l_0} e^{jl_0 \varphi} \quad (5)$$

where $l_0$ and $l_i$ are the carrier mode and modulated modes, respectively. $a_i$ is the modulation index of $l_i$ and $A_{l0}$ are the amplitudes of carrier modes. $\psi(\varphi)$ is the function of the sampled field along the circumference of the $z$-axis. Here, the shape-tailored metasurfaces are proposed to generate the envelope-modulated OAM vortex beams as expected. According to Equation (5) and the shape-related modes discussed in section 2.3, the shape of the designed metasurface could be described as follow:

$$R(\varphi) = r_0 + \sum_i r_i \sin(m_i \varphi) \tag{6}$$

where $r_0$ is the initial radius, $r_i$ is modulated radius, and $R(\varphi)$ is the envelope function of the metasurface. According to the envelope-modulated theory,[59] the shape-tailored metasurfaces could generate modulated OAM modes as well as the carrier mode $l_0$. The modulated OAM modes $l_m$ in the OAM spectrum are given as follow:

$$l_m = l_0 \pm m_i \tag{7}$$

**Figure 5**a,b,c show three specific shape-tailored metasurfaces for verification, where the arrangement region(yellow) of metasurface is defined by Equation (6) and we only place the meta-atoms at the selected region. The corresponding parameters in Equation (6) have been given in specific models and the unspecified values of $m_i$ and $r_i$ default to zero. Here, the envelope configurations are used to produce modulated modes, while the meta-atoms are rotated as usual to produce carrier modes. Under the excitation of the LHCP-plane wave, the corresponding co-polarization $E$-field pattern could be obtained. Figure 5d depicts the far-field numerical results for the model I ($l_0 = -1$, $l_m = -1 \pm 3$) and the specific OAM spectra at the polar angle $\theta = 9°$ and $\theta = 21°$ (where the Fourier transform is implemented along the dotted lines). The Fourier sampling field depicted by the dotted line at $\theta = 9°$ shows a $l = -1$ spiral phase front in $\angle E$ pattern and a three-periodic modulation in $|E|$ pattern, which is just the envelope-modulated OAM vortex beams in Equation (5) as expected. The corresponding spectrum also verifies the generation of the modulated modes $l_m = 2$ and $l_m = -4$. It is worth to mention that the generated modulated modes still satisfy the OAM-related theory, and the higher-order mode ($l_m = -4$) has a larger divergence angle ($\theta = 21°$ in this case).[2] Furthermore, the Fourier sampling field depicted by the dotted line at $\theta = 21°$ in Figure 5d shows more details about the $l = -4$ modulated mode generation. Due to the interference of the reflected field, three $l = -1$ spiral phase periods circled by the dotted line are added into the $l_0 = -1$ spiral phase front in $\angle E$ pattern. Then, a $l = -4$ spiral phase front is generated, and the energy weight of $l = -4$ dominates the OAM spectrum at $\theta = 21°$.

To better illustrate the performance of generated multimode vortex beams, a three-dimensional (3D) OAM spectrum analysis is demonstrated, where all the normalized OAM spectra with different $\theta$ from 0° to 90° are provided. Figure 5e shows the corresponding 3D OAM spectra of Model I at 9, 14 and 19 GHz, where a carrier mode $l_0 = -1$ along with two modulated OAM modes $l_m = -4$ and 2 followed by the Equation (7) are dominating in the spectra as expected and other OAM modes are small enough to be ignored. These results not only prove the feasibility of our proposed envelope-modulated scheme but also show the capability for the broadband design of OAM metasurfaces.

The result of model II in Figure 5f illustrates that this modulation method is not an ideal modulation as that in Equation (5). Some unexpected modes, an obvious crosstalk mode or secondary modulation mode ($l = 3$, i.e. $l_0 + 2m$) can be introduced in this case. Specifically, this non-ideal modulation is due to the limited shape-tailoring capability with different radius, while such a shape-tailoring will also destroy the continuity of the metasurface. These two problems can be well alleviated by reducing the modulation index ($r_1/r_0$) and the specific model III is shown in Figure 5c, where the $r_1$ is reduced from 30 *mm* to 10 *mm* in comparison with model II. Although, a small modulation index ($r_1/r_0$) will result in a small energy weight of modulated mode, the result of the model III in Figure 5g shows that the crosstalk mode ($l = 3$) is almost eliminated. Notice that both the carrier mode and modulated modes can be effectively generated by these three models. The obtained results verify the feasibility of our proposed shape-tailored metasurfaces, and the shape-tailored method offers a new way for OAM generation and OAM-spectrum-tailored technology.

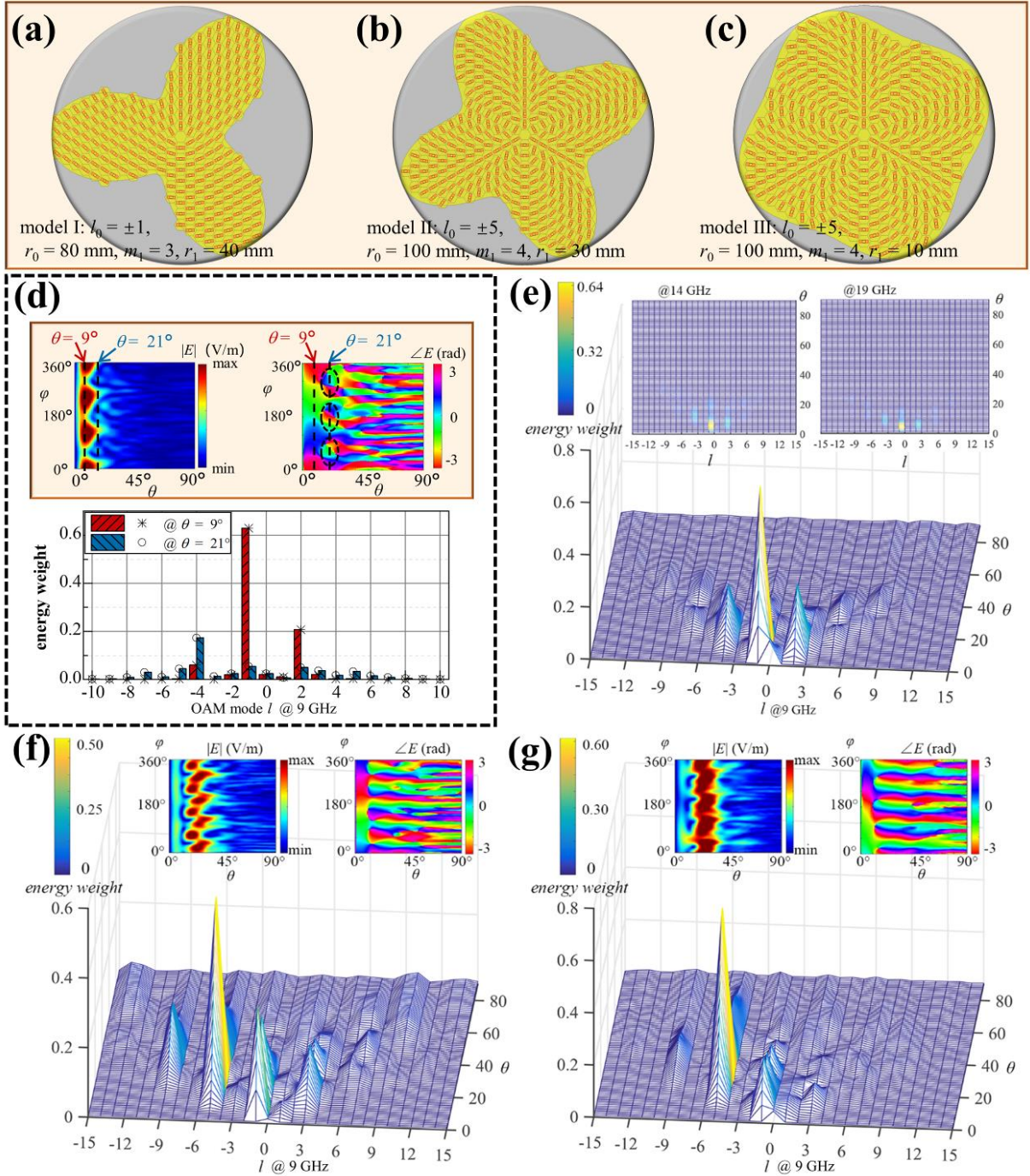

**Figure 5.** Schematics of the generated shape-tailored metasurfaces with specific modulated parameters named by a) model I, b) model II, and c) model III. d) Far-field numerical results and the specific OAM spectra analyses for model I under the excitation of the LHCP-plane

wave at 9 GHz ($l_0 = -1$, $l_m = -1 \pm 3$). The corresponding 3D spectral analyses for e) model I, f) model II, and g) model III.

## 2.5. Generation of comb-like OAM Spectra

The optical frequency comb,[60] defined as a laser spectrum composed of a series of discrete comb-like frequency lines, is a precise tool for measuring different frequencies, building optical atomic clocks.[61] Recently, an OAM comb, composed of equally spaced OAM modes, is generated by pinhole plates based on the famous Young's interference experiment.[46,62] Here, we show another simple way to generate an OAM comb based on the proposed shape-tailored design. For the carrier mode $l_0 = 0$, we can simply use a metal ground to fabricate this "metasurface", and each pair of "tooth" in the OAM comb is controllable by tailoring the metal ground, which provides more flexible control over OAM combs than pinhole plates. **Figure 6**a shows the schematics of the OAM-comb-oriented "metasurface" (actually a metal ground) and the corresponding cross-polarization far-field pattern. Although the energy percentage of carrier mode $l_0 = 0$ in the 3D OAM spectrum occupies most of the energy, the corresponding comb-like OAM spectrum at $\theta = 17°$ with expected OAM mode $l = 0, \pm 3, \pm 6$ illustrates the availability of the shape-tailored metasurface. It is important to mention that there are no rotated meta-atoms placed in the metal ground, which verifies this simple and effective amplitude-modulated strategy.

If the metasurface with rotated meta-atoms is used instead of a metal ground, the spectral center of generated OAM comb could shift to be the value of carrier mode $l_0$. Figure 6b shows a specific case for $l_0 = -1$, where the OAM modes in the OAM comb can be given by $l = l_0, l_0 \pm m_1, l_0 \pm m_2$. As the modulated radius $r_1$ reduces from 40 *mm* to 20 *mm*, the energy percentage of the modulated OAM modes $l_{m1} = 2$ (with energy percentage ≈ 10%) in Figure 6b are lower than that (with energy percentage ≈ 20%) in Figure 5e. The modulated OAM modes $l_{m2}$ (-7 and 5) are also generated as expected. This simple shape-tailored metasurface allows us to manipulate the OAM spectrum elegantly, and particularly each pair of modulated OAM modes $l_m$ in the spectrum are adjustable in terms of their total numbers ($\Sigma i$), amplitudes ($r_i$) and locations ($m_i$).

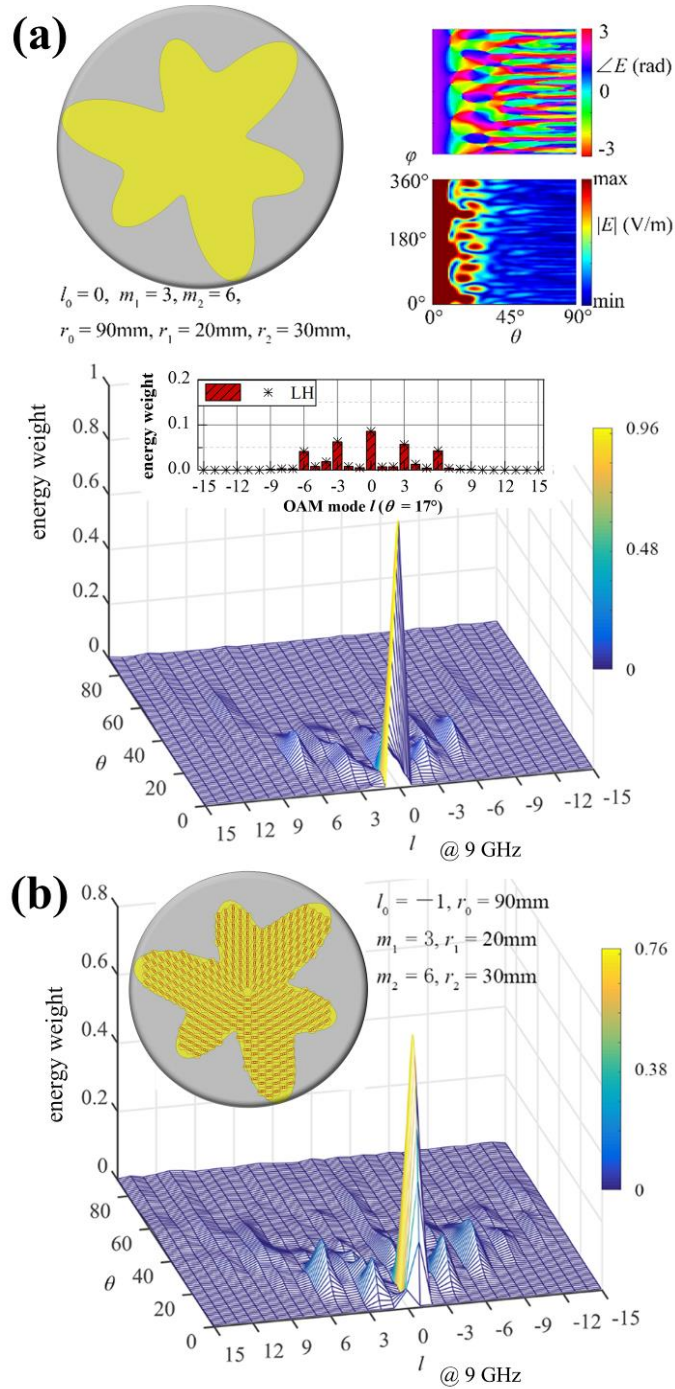

**Figure 6.** Schematics of the shape-tailored metasurfaces and their corresponding 3D spectral analysis under LHCP plane wave excitation at 9 GHz. a) The shape-tailored "metasurface" (metal ground) with carrier mode of $l_0 = 0$, and the corresponding cross-polarization far-field

pattern and its specific spectrum at $\theta = 17°$ are included. b) The shape-tailored metasurface with a carrier mode of $l_0 = -1$ and the corresponding 3D co-polarization spectral analysis.

## 3. Experimental Section

To further verify our proposed method, an experiment is implemented and the setups are shown in **Figure 7**, where a $l = \pm 2$ metasurface and a right-hand circular polarization (RHCP) antenna are fabricated and fixed on a 3D printed bracket with a distance of 200 *mm*; The RHCP antenna is designed by a wideband Archimedes spiral antenna, which can benefit from its small cross-sectional radius of 14 *mm* to ensure negligible influence on the reflected vortex beams. The probe antenna is set 300 *mm* away from the metasurface. The sampling plane size is 400 *mm* × 400 *mm*. To keep the integrity of the metasurface for further modulation, a shape-tailored absorbing material is covered on the top of the $l_0$ metasurface to replace the shape-tailored step. The photograph of the specific experimental setup is also illustrated in Figure 7b, where the specific carrier mode $l_0$ is 2 or 0. The working frequency is 9 GHz.

The experimental sampling results and corresponding 3D spectrum analyses are demonstrated in **Figure 8**. The sampling near-field in Figure 8a shows a 2-period envelope change and a $l_0 = 2$ helical phase wavefront. The 3D OAM spectrum verifies this modulated OAM vortex character. In this case, both carrier mode and modulated mode can be detected from a wide range of sampling radius $r$ from 50 *mm* to 200 *mm*. In contrast, some crosstalk patterns (if any) only appear in a small range of $r$. Figure 8b shows the metal ground case ($l_0 = 0$). Although the sampling field in Figure 8b does not show the energy null and helical phase wavefront (the character of vortex beams), the modulated *E*-field envelope generates the OAM modes $l_m = \pm 2$. Hence, the obtained experimental results not only verify the proposed modulated scheme works well at near-fields but also provide another feasible method to generate modulated OAM beams ($l_0$ metasurface covered by shape-tailored absorbing material).

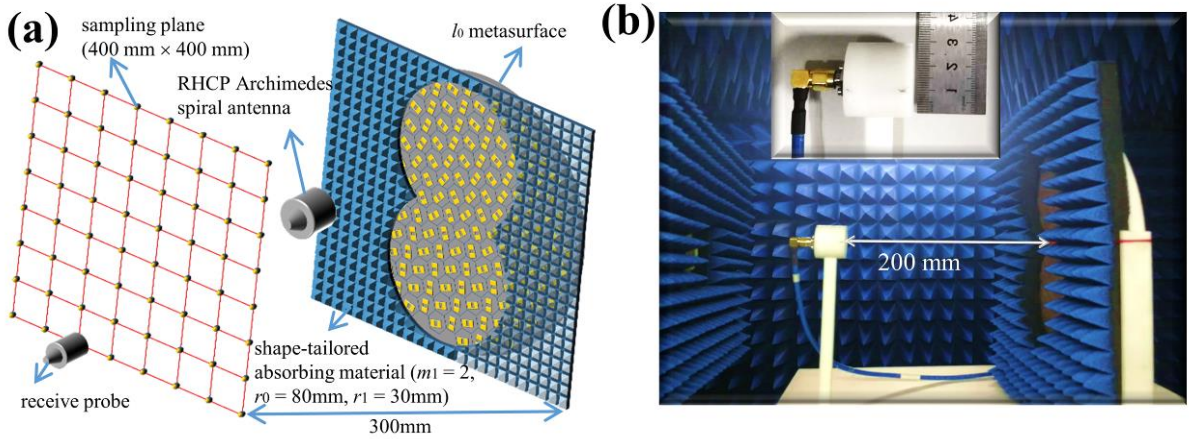

**Figure 7.** a) The experimental setup of near-field scanning to measure the near-field electric field. b) The photograph of the specific experimental setup.

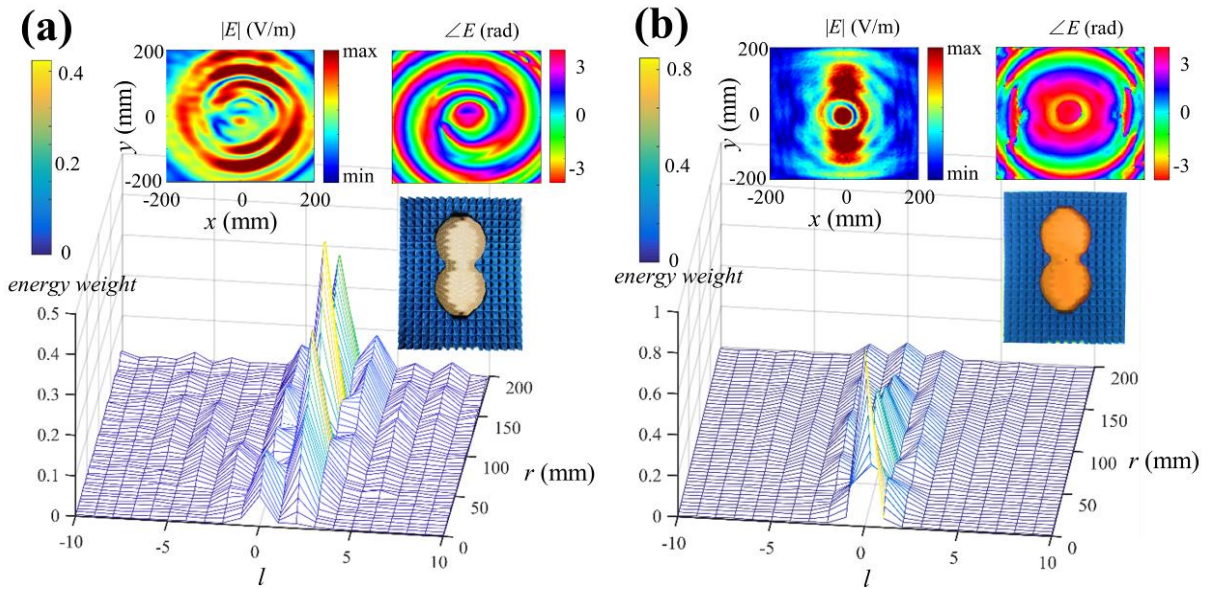

**Figure 8.** The near-field scanning results and corresponding 3D spectrum under the excitation of the RHCP antenna at 9 GHz. a) Co-polarization field generated by the covered $l_0 = \pm 2$ metasurface. b) Cross-polarization field generated by the covered $l_0 = 0$ metal ground.

## 4. Conclusion

In this work, a simple but efficient scheme for the modulated OAM spectrum generation by shape-tailored metasurface has been proposed, implemented, and experimentally verified. Duo to the wave interference, the modulated amplitude of the reflection field can be obtained by the tailored metasurface. Several modulated OAM spectra and the comb-like OAM spectra have been generated by the well-defined shape-tailored metasurfaces. As a superposition of specific OAM modes, the generated spectra would be useful for various OAM-based applications. Moreover, the proposed shape-tailored method with the triangular lattice is well suitable for other metasurfaces without affecting their bandwidths and response, which provides an avenue for multi-dimensional high-performance broadband light control. Generally, although this work focuses on OAM-based applications, the proposed method increases the degree of freedom of EM wave control, and also be easily transferred to other amplitude-based devices. In addition, it is not limited to the radio-band reflection metasurfaces, which could be naturally extended to optical, transmission, and other programmable metasurfaces.

**Supporting Information**
Supporting Information is available from the Wiley Online Library or from the author.

**Acknowledgements**
This work was supported in part by the National Natural Science Foundation of China under Grant 61971115, Grant 61975177, Grant 61721001, Grant 61622106, and in part by Sichuan Science and Technology Program under Grant 2018RZ0142.

**Conflict of Interest**
The authors declare no conflict of interest.

Received: ((will be filled in by the editorial staff))
Revised: ((will be filled in by the editorial staff))
Published online: ((will be filled in by the editorial staff))

**Table of contents:**

**Through tailoring the shape of metasurfaces,** the power in orbital angular momentum (OAM) spectrum can be modulated and redistributed. A universal modulation relation is established between the spatial arrangement of metasurfaces and the OAM spectra. Modulated OAM spectra and comb-like OAM spectra are both theoretically and experimentally demonstrated, which provides an avenue for multi-dimensional high-performance control of electromagnetic waves.

**Keywords:**
OAM spectrum, orbital angular momentum (OAM), Pancharatnam-Berry phase, shape-tailored metasurface, triangular lattice

*L.-J. Yang, S. Sun\*, W. E. I. Sha\**

Title:
**Manipulation of Orbital Angular Momentum Spectrum Using Shape-Tailored Metasurfaces**

ToC figure:

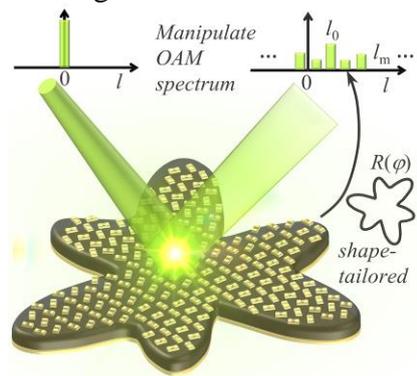